# EVOCATION AND ELABORATION OF SOLUTIONS: DIFFERENT TYPES OF PROBLEM-SOLVING ACTIONS

## AN EMPIRICAL STUDY ON THE DESIGN OF AN AEROSPACE ARTIFACT[*]

# EVOCATION ET ELABORATION DE SOLUTIONS: DIFFERENTS TYPES D'ACTIONS DE RESOLUTION DE PROBLEME

## UNE ETUDE EMPIRIQUE DE LA CONCEPTION D'UNE STRUCTURE AEROSPATIALE

Willemien VISSER[**]

**ABSTRACT**. An observational study was conducted on a professional designer working on a design project in aerospace industry. The protocol data were analyzed in order to gain insight into the actions the designer used for the development of a solution to the corresponding problem. Different processes are described: from the "simple" evocation of a solution existing in memory, to the elaboration of a "new" solution out of mnesic entities without any clear link to the current problem. Control is addressed in so far as it concerns the priority among the different types of development processes: the progression from evocation of a "standard" solution to elaboration of a "new" solution is supposed to correspond to the resulting order, that is, the one in which the designer's activity proceeds.
Short discussions of • the double status of "problem" and "solution," • the problem/solution knowledge units in memory and their access, and • the different abstraction levels on which problem and solution representations are developed, are illustrated by the results.

**RESUME**. Des observations ont été conduites auprès d'un concepteur professionnel, pendant son travail sur un projet de conception dans l'industrie aérospatiale. Les données du protocole ont été analysées pour identifier les actions que le concepteur utilise pour développer une solution au problème posé. Différents processus sont décrits: de la "simple" évocation d'une solution qui pré-existe en mémoire jusqu'à l'élaboration d'une "nouvelle" solution à partir d'unités mnésiques sans aucun lien évident avec le problème en question. Le contrôle est abordé pour autant qu'il intervient dans la détermination de la priorité parmi les différents types de processus de développement de solution: on suppose que la progression qui va de l'évocation d'une solution "standard" à l'élaboration d'une "nouvelle" solution correspond à l'ordre qui en résulte, c'est-à-dire, l'ordre dans lequel l'activité du concepteur procède.
De brèves discussions • du double statut de "problème" et de "solution," • des unités de connaissances concernant les problèmes et leurs solution(s) en mémoire et l'accès à ces connaissances, et • des différents niveaux d'abstraction auxquels des représentations de problème et de solution sont développées, reçoivent ensuite des illustrations dans la présentation des résultats.

**Mots clés** : sciences cognitives, psychologie cognitive, résolution de problème, conception, étude de protocole, représentation de problème, accès aux connaissances, activation

---

The study presented in this paper analyzes problem-solving involved in design of artefacts, a domain which has received less attention than software design (see Visser 90a and 90b for references to other design studies). The activity is examined at a great level of detail, and analysis focuses on the different types of processes involved in solution development. Whereas empirical research on design mostly concerned the evocation of knowledge, this study additionally analyzes solution elaboration, the other main development mode.

The introduction provides a general presentation of the design process and a discussion of some aspects of problem-solving which are central in the paper. Section 1 presents the method used to collect data on the design activity. Section 2 uses these data to describe the different types of solution-development actions. The last section discusses these results and provides suggestions for future research.

**INTRODUCTION**

At a high level of abstraction, the design process may be described as
• proceeding in iterative cycles, and
• leading progressively from a global problem specification to a detailed solution.

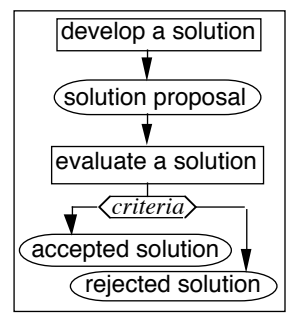

In each cycle, a solution to a problem is developed and then evaluated. Development of a solution leads to a solution proposal. Each proposed solution is evaluated and may be accepted or rejected (possible finer distinctions are not used in this text) (see the fig. 1).

In order to attain a final, implementable solution, problems are worked out in two directions. Both types proceed at successive levels, reducing abstraction until the implementation level is reached. On the one hand, problems are worked out in breadth, decomposed into sub-problems, which imposes integration of intermediate solutions. On the other hand, problems are worked out in depth: each solution to a problem is taken as a problem to be solved until a final solution is reached. This solution has to be detailed and concrete enough to specify its manufacturing (that is, in the case of artefacts' design).

figure 1: A problem-solving cycle

This description however does not characterize the actual design activity, which has rather an opportunistic organization (see Visser 90b, for a survey of relevant psychological research in the domain of software design). In an empirical study on specification, (Visser 90a) addressed the control level by posing the question: which mental processes lead the actual activity to deviate from a neat, hierarchical structure? This paper is going to analyze design at the level of the individual design actions.

Before presentation of the study itself, a discussion will introduce four aspects of problem-solving which underlie the approach taken in the study.

Problem or Solution: A double status. Design problem-solving starts with a *problem*, that is, specifications provided by a client to a designer which specify more or less precisely the artefact to be designed, and it has to lead to a *solution*, that is, specifications provided by the designer to the workshop which specify precisely how to manufacture the artefact.
The problem-solving path, consisting in a transformation from the initial problem specification into a solution, comprises a great number of intermediary states, each with a double status. Until the final implementation level is attained, each solution developed for a problem constitutes itself a new problem to be solved. That is why, in this text, the same entity may be referred to as a "solution" or as a "problem," depending on whether it is an output or an input of a problem-solving action.

Problem/Solution knowledge: Structure and access. A knowledge unit may constitute a "solution" to a "problem" without being coded as such in memory. For example, "63" is the solution to the problem "How much is 47+16?," but for only a select few the mnesic entity corresponding to 63 may have an attribute such as "result of the addition of -: v1=47, v2=16." The majority of people probably also do not have a knowledge unit such as "47+16=63" (whereas, "2+2=4" or "3+3=6" probably are stored as such in many peoples' memory, for example, as values of the attribute "example" of the schema of "addition").
Nevertheless, people probably do have units in memory integrating knowledge about classes of problems and their solutions (see Chi 81). Adopting the conceptualization in terms of "schemata," this type of knowledge will be referred to as "problem/solution schemata" with "problem attributes" and "solution attributes" (although it is not evident if these are different types of attributes).

However, a great part of the knowledge a person possesses on problems and their solution(s) cannot - or can not yet- be considered to be schematic: it does not concern a class of problem situations, but a particular problem and its solution(s). In this text, except if there is evidence for problem/solution knowledge being schematic, it will be referred to as "problem/solution associations" (or, in an abbreviated form, "problem/solutions").

Schemas and problem/solution associations are similar in that, next to a "label" identifying them, both have attributes. But they are different in that a schema's attributes are variables, whereas a problem/solution attribute has a particular value.

In various models of cognition (see, for example, Anderson 83), a *spreading activation* mechanism is supposed to underlie the access to this knowledge. This assumption is adopted in the present study.

<u>Scope of a solution's activation</u>. When a problem/solution is activated, what is activated? It is certain that its label is activated, but the question remains whether or not all its attributes and their values are activated. The data at disposal do not allow us to decide.

We do not take a position on the question of which attributes and values are activated, but formulate a weak hypothesis: among the attributes which are activated on problem/solution activation, one can distinguish at least two types: those which are evaluated when "the solution" is evaluated (the solution's evaluation amounts to their evaluation), and those which are activated but not included in this first evaluation of the solution.

An example of problem/solutions will be presented in a later section, when the design problem used for illustration will have been described (see the fig. 4).

<u>Different forms of problems and solutions: concept - function - materialization</u>. In engineering problems, the client's specifications generally specify the artefact to be designed in terms of a goal to be attained under certain functional and material constraints: an automatic installation for manufacturing certain parts at a certain speed, a satellite fulfilling a certain mission with a certain useful charge.

Even if the final solution of the design process consists of the specifications of an artefact -that is, the solution is not an artefact itself- it is considered to be a *material* solution: it specifies *how* to manufacture, and with which *physical components*. Before this final material solution is proposed, the global design problem-solving process often proceeds by analyzing and further specifying the problem specifications, in order to transform the goal in terms of a *concept* or a *function* which allows attainment of this goal.

On the level of intermediate sub-problem solving, the same progression may be observed, that is, from a goal, via a concept or a function, to its materialization. At this level however, a problem, defined in terms of a goal to be attained, may be solved immediately in material terms -a situation never observed at the global design process level, except in the case of routine design.

Examples of these different problem and solution forms will be given in the two next sections.

## 1. AN EMPIRICAL STUDY OF THE DESIGN PROCESS: METHOD

The data have been collected in an observational study conducted on a professional designer working on a real design problem in an aerospace factory.

### 1.1 Procedure: observation & simultaneous verbalization

The designer, working full time on a project, has been observed during five weeks, at the rate of 3-4 days a week. Observation occurred in the middle of the design project.
The designer's normal daily activities on the project were observed without any intervention, other than to ask the designer to verbalize, as much as possible, his thoughts during problem-solving.

### 1.2 Participant: a professional, experienced designer

The observed designer (D) had some 30 years of professional experience in the Research and Development division of the aerospace factory. For the past four or five years, he has been involved in designing aerospace structures similar to those being studied here (a new type of antenna).

### 1.3 The observed task: design of an unfurling antenna

The global problem to be solved by D was the design of an unfurling antenna (according to particular specifications). Two sub-problems (each with their sub-problems etc.) have mainly been handled during the four to five weeks that observations were conducted: the "upholding system" problem and the "vertical tubes" problem. These problems will be used to illustrate the different solution-development processes.

The unfurling antenna. The antenna to be designed is a *model* on which to study the unfurling mechanism (the design of which is part of the current project). The entire antenna does not have to be manufactured, but only one of its component unities (called nevertheless "the antenna" in this text). Right from the start, D adopts the attributes "form" and "type of reflecting surface" which had been used on a previous unfurling antenna. The antenna's form will be a polyhedron, consisting of triangles at the top and base, and three rectangles as the remaining faces. The edges are tubes: the six horizontal tubes are jointed (allowing the antenna to unfurl), the three vertical tubes are stiff. The antenna's reflecting surface will be a supple tissue of a particular material (see the fig. 2).

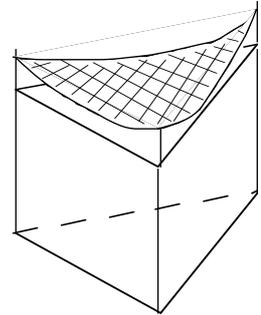

figure 2: The antenna: its form and its reflecting surface

The "upholding system" problem. In an analysis of a previous unfurling antenna, D identified a problem with the antenna's unfurling mode. The analysis led him to identify the risk that, during the antenna's furling, the reflecting surface would tangle. In order to avoid this, a system was proposed for upholding the reflecting surface above the level of the vertical tubes. This upholding system had two subsystems: a *support system* -holding the tissue above the level of the tubes- and a *pulling system* -a system of pulling elements tightening the support system by their pulling action.

The "vertical tubes" problem. Solving the "upholding system" problem on the current project, especially specifying its material solution, has led D to identify a second problem: the need to adapt the vertical tubes to the upholding system. How to do this needs to be solved.

The next section presents the main solution-development modes identified in the data gathered on the antenna-design project (see Visser 90a, for the approach taken to data analysis).

## 2. DIFFERENT SOLUTION EVOCATION & ELABORATION PROCESSES

Different types of processes may contribute to solution development: from the "simple" evocation of solutions existing in memory, to the elaboration of a "new" solution out of mnesic entities without any clear link to the current problem.

Confronted with a problem to be solved, a designer is supposed to start solution development by attempting to evoke a ready-made solution from memory, that is, to match a problem/solution association or schema. Only if this fails, elaboration of a solution is engaged (see the fig. 3).

The global distinction between "evocation" and "elaboration" is not an absolute one. Except when its basic solution constituents come from an external information source, the elaboration of a solution uses mnesic entities, which will have been activated, by evocation (that is, approximately, "unguided" activation), or by "guided" activation (see below).

The different types of problem-solving actions will be presented as particularizations of these two modes.

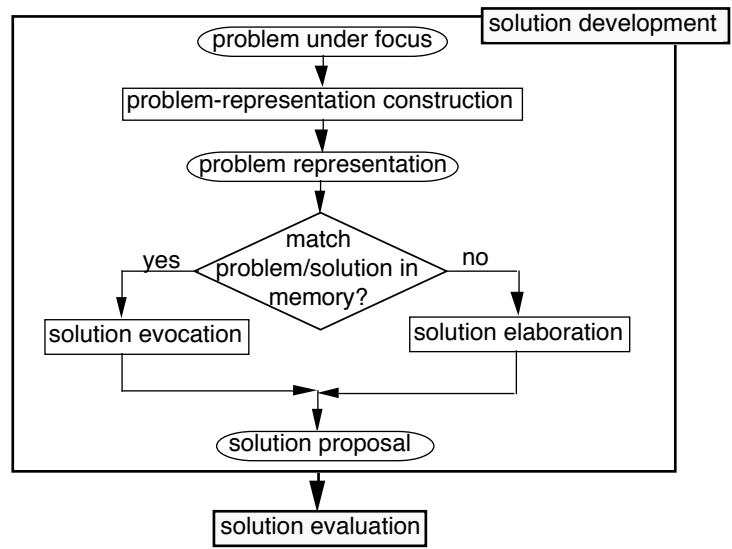

figure 3. Solution development: evocation or elaboration

Their presentation will proceed by general descriptions illustrated by observations on the solving of the upholding-system and the tubes problem.

## 2.1 Evocation of a solution

Having a problem to solve, a person starts with the construction of a representation of the problem (a process not detailed here; see Richard 85). The mnesic entities of the resulting representation will be the "sources" for activating other mnesic entities. If they match the problem part of a problem/solution or (some of) the problem-attributes of a problem/solution schema, the corresponding solution is evoked.

If only a solution stored as such may be evoked, coming up with a stored solution may require more than the activation involved in evocation, which is "unguided," or rather, "only" guided by the problem representation. Activation of an analogous solution (that is, the solution to an analogous problem), for example, first requires generation of the appropriate sources (see below).

Different levels of experience may mean that one designer knows of -that is, has stored in memory- more solutions for design problems than another. Solutions a designer knows of may have been developed in the past by himself or he may know of their existence from colleagues or from literature. So, an experienced designer may evoke solutions which a less experienced colleague would need to elaborate.
Experience -or "expertise" (see Kolodner 83)- also introduces differences in the internal structure of knowledge, its organization and its use (see also Chi 81). So, if confronted with a problem, two designers which differ in expertise in the problem domain both evoke a solution, one may expect them to evoke sometimes different solutions, because • they may use different problem representations, and • the same representations may activate different solutions.

Different types of stored solutions may be distinguished, according to their readiness to be activated (discussed later on), and to their "historical link" to the current problem.
Solutions acquired by experience refer to solutions which exist in memory before the current global-problem solving. These "pre-existing" solutions (standard" and "alternative") will be further discussed below. But the current problem-solving also leads to solutions being stored in memory. Some may be explicitly "shelved," that is, • developed, proposed, and not adopted, but • put aside "for possibly later usage." Others will enter long-term memory, even if the designer "did nothing" -consciously- to retain them. Their storage may depend on several factors: the length of time during which the solution has been under focus and the type of processing it was subject to, for example, the time and effort required to elaborate the solution, the number (and perhaps type) of evaluations it was subject to, the time during which it was accepted.

<u>Determine the type of solution development and the type of solution</u>. Settling, with respect to a solution whose proposal is observed in the protocol, if it has been evoked or elaborated is not always evident. And if one has decided, for certain reasons, that a solution has been evoked, the next distinction to be made is between the pre-existing ones and those developed on the current project, and, among this last group of solutions, between the "shelved" ones and the others.

* Example of hesitations on this point and of arguments used to decide on a type of solution (development). Confronted with an interface problem, D proposes a solution, saying "Here I may use the roulette wheels." Does he evoke or elaborate this solution? Elements in the protocol pleading for evocation are "here" and "the (roulette wheels)."
Next, there are elements in -and external to- the protocol possibly pleading for the solution being one developed on the current project. Two problem-solving cycles earlier, a roulette-wheels solution had been proposed for another interface problem. Negative aspects had been identified on the solution, but it was not explicitly rejected: D rejected the solution it was part of.
When, afterwards, the experimenter asked him, D said that he had not used this solution on any previous project.

<u>The "standard" solution and "alternative" solutions</u>. Design problems generally have more than one solution. This does not mean that if a designer knows of a solution to a problem, he knows several. But if he does, one among them is activated more strongly on evocation, and so evoked first ("the" standard solution). If the designer possesses alternative solutions to this standard solution, they may be supposed to be evoked -one at a time- in the next development cycle(s), if the standard solution is not accepted after evaluation.
The standard solution (to a particular problem) may be
• if there is only one, the solution the designer has developed -and adopted- in the past, or
• if there are several ones, the solution he prefers, or judges to be the "best" (for being the simplest or the cheapest, for example).
For the upholding and tubes problems and their sub-problems, there are

**upholding system**
<u>function</u>
...
<u>components</u>
   support system &
   pulling system
...

several examples of standard solutions -and of their evocation. There is no example of alternative pre-existing solutions -and thus nor examples of their evocation. An explanation may be that D possesses them, but did not need to use them, because their corresponding standard solutions were not rejected. This is possible -D evoked standard solutions which he did not reject-, but there are also several examples of standard solutions which were rejected.

A question which would have needed an answer, before making the preceding claim, is: When does one consider a solution to be rejected? If at least one of its attributes and their values is rejected? The answer depends -also- on the answer to the question of the "Scope of a solution's activation" (see above).

* Example of an evoked standard solution which is never rejected. The value of the attribute "diameter" of the tubes problem/solution is never rejected. It keeps the value evoked at its initial development: Ø=30mm.

The following example not only illustrates the questions tackled here, but is also the example referred to in the "Scope" paragraph.

* Example of an evoked standard solution which is adopted, but several attribute values of which are rejected on later evaluation. The upholding-system solution proposed and adopted at the start of the project is a standard solution. It had been used on a previous unfurling antenna project. It seems to bring along at least one attribute ("components"), with its value: "pulling system & support system." "Pulling system" and "support system" may be considered each to be problem/solution associations with their attributes and their values, some of which seem have been brought along with the proposal of the upholding system solution.
The value (v) of the upholding-system attribute "decomposition" is never rejected: v="support system & pulling system." However, when focus is on the pulling system, three of its attributes' values are evaluated and rejected: "starting point of the pulling system's elements: v = center," "instance of force: v = workshop operator" and "nature of pulling elements: v = wires" (see the fig. 4).

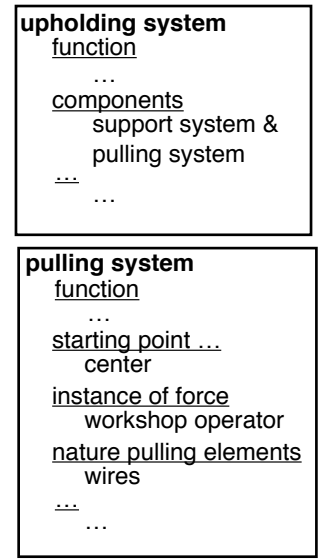

figure 4: Two examples of pre-existing problem/solutions
key: in **bold**: **label**
     underlined: attribute
     standard: value

There is an example of an alternative solution on the global antenna project. It is a solution D possesses in memory, but does not need to evoke, because the standard solution is not rejected. This alternative solution is identified by the experimenter, when she asks D if the solution proposed is the only one.
One argument taken as evidence for evocation rather than elaboration is the immediateness of the answer. Another argument is a colleagues citing of the same solution (one adding that it is the solution used by a competing factory).

* Example. Alternative solution for "geometry of antenna unities:" "triangular faces, base and top" (the standard solution, evoked on the current project, is "rectangular faces and triangular base and top;" see the fig. 2).

## 2.2 Elaboration of a solution

Elaboration may take several forms. A distinction may be made according to the character of the material which constitutes the starting point for elaboration. This may be • the representation of a problem which has not yet received a solution during the current problem-solving (elaboration of a "new" solution) or • a solution previously developed and proposed, but rejected (elaboration of an alternative solution). The rejected solution may be considered to function as a problem which conveys, next to the constraints already taken into account in the previously proposed solution, supplementary constraints introduced by the solution's evaluation (those underlying its rejection). Presentation will start by the second mode, which is supposed to involve a "simpler" development than the first.

**2.2.1 Elaboration of an "alternative" solution.** For many rather specific problems -such as the problem of the starting point of the pulling system's elements in a downward direction- D knows of only one solution, whose evocation leads to its proposal. Afterwards, if he rejects it, but he has no other solution in memory, a solution has to be elaborated.

<u>Generalizing the problem</u>. This strategy consists in constructing a more general representation of the problem than the previous one, in the hope that the new representation will activate a new problem/solution (schema). Generalizing amounts to setting aside some attributes and focusing on others. How this selection is made remains open to question.

* Example. D rejects the solution for "starting point of the pulling system's elements for their pulling in downward direction from the support plane," which is "center." Using his knowledge that the support plane is a triangle, D generalizes this particular starting-point problem to the problem of "points of a triangle," under the constraint: v≠"center" . He comes up with "the angles" and then proposes this value as a new solution to his starting-point problem.

<u>Solution modification heuristics</u>. If the new problem representation does not lead to a solution being evoked, a heuristic may be used for modifying the rejected value. An example is: "invert the value."

* The example presented has been interpreted as exemplifying this heuristic. This may be criticized. D rejects the solution for "path of pulling elements with respect to the vertical tubes, which is "through (the tubes)." The new solution which he proposes is "along (the tubes)."

<u>Going up and then down again in the solution abstraction hierarchy</u>. Generalizing may consist also in reconsidering the problem at a higher abstraction level. This implies a solution development in at least two steps: elaboration, first, of an alternative solution at a higher abstraction level, and then of a more specific solution to the problem constituted by the abstract solution.

* Example. The evaluation of the solution "instance of force: v = workshop operator" has not only led to the introduction of the constraint "v ≠ workshop operator," but has put also a second constraint on the new solution: "the instance must be practical." D neither has elements to evoke, nor to elaborate, an alternative-to-the-workshop-operator solution: he does not possess a category of "practical instances," which would allow him to evoke a solution; neither has he a category of "instances of force" (in the context of pulling actions) which could provide him with an alternative to the workshop operator solution.
This may explain why -whereas the rejected solution was already a material one- D first comes up with a new solution at the conceptual level: "the instance must be without human intervention." He then materializes this solution into "instance of force: v = a weight."

<u>Adopting a default value</u>. This strategy applies as well to the elaboration of an alternative solution, as to the elaboration of the "first" solution to a problem.

* Example of a first solution elaboration. Focussing on the initial roulette-wheels solution (see above), the first attribute D selects to specify is "positioning (of the roulettes)." D does not have in memory a standard solution to the problem of how to position several roulettes at the entrance of a tube. He seems to generalize the problem to the one of "positioning objects in a plane" and adopts, what may be considered to be the default value, "adjacent."

**2.2.2 Elaboration of a "new" solution.** Elaboration of an alternative solution is a solution-development mode that is only applicable when one has already developed a solution (the evaluation of which has led to rejection). This solution and the result of the evaluation constrain the alternative-solution elaboration. But without such a "reference," and without a solution in memory, the solution has to be elaborated from scratch.
If the global design process -even of a completely "new" artefact- never proceeds from scratch, solving of sub-problems may. "Elaboration from scratch" remains to be defined. Here it is considered to apply to solutions which are constructed from material which is only analogous, or even without any clear link to known, past analogous designs. This solution-development mode is probably the most difficult for the designer.
Various processes and strategies may be supposed to be used. Those presented below were observed for the problems studied in this paper, or elsewhere on the antenna project. Some remarks will be made with respect to others which have not been observed in the study, but are known from the research literature.
*Problem decomposition*. Empirical software design studies present problem decomposition as a very important -if not the main- strategy in expert design. By definition, a design problem is decomposed, in that it is transformed into other problems to be solved, but -as used in the design studies literature- the concept conveys many presuppositions, especially: resulting from "planning" and resulting in a "balanced" solution. In (Visser 90b) some critical remarks are formulated with regard to this supposed status of decomposition in design problem-solving, but the question would require a detailed discussion.

* Example. As noted already, the upholding system solution was defined as having two components, a pulling system and a support system. These partial solutions did not result from an action of decomposing the upholding system problem, but from evoking an existing solution.

*Deductive reasoning*. This form of problem-solving has often been studied in problem-solving research, but seldom in the context of "real-world" tasks, such as industrial design, for example. Although it surely occurs in these "natural" conditions, it was not observed on the antenna project.

Brainstorming. The term "brainstorming" is proposed in this paper to refer to a very "weak" "search" mechanism: transitions from one mnesic entity to another are not a function of rules (as in deductive reasoning), or structural or surface similarity (as in analogical reasoning). Search is only "guided" by the nature of the mnesic "source" entities which are firing and by the links existing between them and other mnesic entities. The sources are firing because they have been processed by D in his construction of the problem representation: that is what makes it possible for a person to "guide" his search. The "idea" D comes up with, or the solution he "thinks about," is the newly activated entity which exceeds a threshold. This may be, for example, because it has been activated by various sources (see example below), but more than one entity may exceed the threshold, and other factors may contribute: these points are not discussed here. Brainstorming thus is the more or less conscious use of the unconscious, "natural" access mechanism, that is, spreading activation.
Not only can brainstorming not be "observed" behaviorally, neither are there elements in the protocol for translating it, other than the designer hesitating, falling silent, and/or coming up with the resulting "idea" or "solution proposal" after a more or less long silence.
The observer may suppose that brainstorming is used when, in the absence of evidence for "stronger" forms of solution elaboration, she/he has a hypothesis with regard to the transition path from one or more sources to the solution proposal. Like the other explanations proposed, such hypotheses remain of course to be verified.

* Example. The evaluation of the solution "nature of pulling elements: v = wires" has led to the identification of several "constraint satisfying," several "positive," and several "negative" aspects of this solution. The solution is not rejected, but shelved, and D tries to develop a solution which • satisfies the constraints on the elements (they must be able to pull and to stretch), • has the positive, • but not the negative aspects of the wires solution (the elements are to be light, but not subject to the risk of tangling when using wires, because of their combined supple- & thinness). First D comes up with the concept of "tape measure," and then he proposes the solution "nature of pulling elements: v = strips" (possibly by analogical reasoning).
The main sources which are supposed to fire and to lead to the activation of "tape measure" are: "light," "stretch," "pull," and a temporary mnesic entity translating the idea of "avoiding the risk of tangling, as exists when using wires, due to their combined supple- & thinness."

With the reservations cited above, the problem-solving path for the problems examined in this study may be considered to show a frequent use of brainstorming.

Analogical reasoning. This process may be supposed to start with the same mechanism as brainstorming. But after having accessed an "analogous" problem/solution, the famous "mapping" has to be made, in order to use the knowledge accessed (not detailed here, see Vosniadou 89). Although this process may be supposed to be used all along the design process, its observation on the antenna project occurred especially in the conceptual design-solution development stage (but see also the example, given above, of D coming up with the material "strips" solution, after having proposed the conceptual "tape measure" solution).

* Example 1. Having adopted an upholding system composed of a support system and a pulling system, the development of a concept for the support system involved an important part of analogical reasoning. D, as well as several of his colleagues, formulated conceptual solutions in terms of familiar systems from other domains: "curtain rails,""train rails" and "railway catenaries" were proposed, for example.
* Example 2. Developing, in a discussion with colleagues, "unfurling principles" for future antennas, D and other designers proposed conceptual solutions such as "umbrella," and other "folding" objects, such as "folding photo screen," "folding butterfly net," and "folding sun hat." The designers thought of taking apart the objects in order to find (possible ideas for) the corresponding material solutions.
Note that all this "folding" objects satisfied the constraint of having a trigger mechanism leading to a complete unfurling of the object which then constituted a rigid surface.

Two examples have been presented, but there are, at least, six others of the same type, that is, examples of a problem (target) for which one or more conceptual solutions are proposed coming from a different domain where the sources bear no surface similarity with the targets. This result seems different of the main results of research on access to analogous source material in memory which found surface similarity to be the dominant factor, producing more sources to be retrieved than higher-order structural (relational) commonalties (see Gentner 89). An explanation of this difference would require a detailed comparison between the situations in which data have been collected.

Simulation is often used for evaluation, but it may also serve solution development (see Visser 90b). On the antenna project it was observed to be used by D when materializing a conceptual solution, or further specifying a material solution (to the evaluation of which the process also contributed). Concrete simulation was often observed to support mental simulation.

* Example. Several versions of a material solution to the "curtain rails" problem (see above) were elaborated. D sometimes made a rough sketch of the top of the antenna, and estimated the required path of the pulling elements, by simulating their downward direction with hand movements. On other occasions, he combined paper cut-outs, wires and pencils (for the tubes), which he moved around the drawing by hand.

## DISCUSSION

This study addressed the activity of artefacts design and focused on the processes underlying individual design-problem-solving actions (rather than on their organization: only the priority assignment among the different types of development processes was described). Having characterized problem-solving as proceeding by solution development and by solution evaluation, this text then concentrated on the first type of activity. Solution development was analyzed as starting by an attempt to evoke a solution; if the designer has no problem/solution in memory (a schema or an association between a particular problem and its solution(s)), a solution has to be elaborated. Evocation may give access to different types of solutions in memory. Different types of elaboration processes were distinguished. Several hypotheses which were formulated to this respect require further confirmation. Examples are the nature and role of "brainstorming" and the factors determining access to analogous material in memory. The data collected on these processes and the proposed interpretations provide directions for such research.

Next steps towards specifying a model of the design activity would be, first, to examine if the results obtained for the control level in our study on specification in the domain of factory automation cells (see Visser 90a) hold for other design tasks, such as the one studied here or such as software design, and, then, to integrate the control and execution levels.
With respect to this integration, next to the classical control topics (main organization of the activity, criteria involved in the choice of the next action, for example; see Visser 90a), other questions, more specific to problem-solving (or even to design), are suggested by the data of this study. Only two will be shortly approached here.
*Selection of problem attributes.* This important control process presupposes the existence of the elements among which the selection is made. Observation, however, suggested that it often goes together with the identification of these elements. Once again, evidence is not simple: in many cases we did not observe that the elements were identified (before a selection among them occurred), but one can suppose that, implicitly, they were. This may be the case, for example, because they made up part of the activated solution (see above the discussion on Scope of a solution's activation).
*Integration of several solutions.* The final solution to a design project integrates a great number of solutions to sub-problems. The type of processing involved in integration probably depends on the moment in the design process when this integration takes place. The designer does not wait until he has developed all component solutions, before integrating them. The data on the antenna design suggest the hypothesis that, confronted with two interacting problems, • problem one is specified rather precisely (the most "constraining" one), • problem two is only minimally specified and • finally, the specification of their interface is started.

* Example. Having specified several attributes of the pulling system, and then having determined the value of one attribute of the tubes, D starts to define their interface. The evaluation of the proposed interface solution (I) leads to modification of both the pulling system (P) and tubes (T) solutions. The next problem-solving actions then consecutively focus on: I, T, I, P, I.

This paper concerned design problem-solving, but the results with respect to the solution development processes may be relevant for other types of problem-solving activities. This holds also - perhaps even more- for the questions formulated, especially those concerning the Scope of a solution's activation. Elements for analyzing this question may possibly be found in the literature on categorization. The results will constrain a general model of cognition.

Acknowledgements. Thanks are due to Françoise Détienne for helpful comments on the content of this manuscript, and to Steve Gibbons for corrections of its form.